\documentclass[letterpaper]{article} 
\usepackage[preprint]{aaai2027}  
\usepackage[hyphens]{url}  
\usepackage{graphicx} 
\usepackage{natbib}  
\usepackage{caption} 
\usepackage{algorithm}
\usepackage{algorithmic}
\usepackage{amssymb}
\usepackage{amsmath}
\usepackage{multirow}

\usepackage{newfloat}
\usepackage{listings}
\DeclareCaptionStyle{ruled}{labelfont=normalfont,labelsep=colon,strut=off} 
\floatstyle{ruled}
\newfloat{listing}{tb}{lst}{}
\floatname{listing}{Listing}

\usepackage{booktabs}

\title{Experience-Calibrated Contrastive Decoding \\ for Mitigating Hallucinations in LM-Based Text-to-Speech}
\author{
    Chenlin Liu \textsuperscript{\rm 1},
    Minghui Fang \textsuperscript{\rm 2},
    Zhonghao Bi \textsuperscript{\rm 1},
    Zekai Su \textsuperscript{\rm 1},
    Rong Wang \textsuperscript{\rm 1},
    Jiqing Han \textsuperscript{\rm 1}
}
\affiliations{

    \textsuperscript{\rm 1} Harbin Institute of Technology, China \\
    \textsuperscript{\rm 2} Zhejiang University, China \\
    chenlin.liu@stu.hit.edu.cn \quad jqhan@hit.edu.cn
}

\begin{document}

\maketitle

\begin{abstract}

Language model-based text-to-speech (LM-based TTS) remains vulnerable to speech hallucinations that deviate from the target text. Existing mitigation mainly relies on architectural changes or additional training, while decoding-time control remains underexplored. We present a conditional information view that distinguishes text-derived \textbf{alignment information} from \textbf{experience information} supplied by acoustic context and learned speech regularities. We hypothesize that an important class of hallucinations begins when alignment support is insufficiently reflected in the selected token at a vulnerable transition. Using predictions from the same speech LM with and without text conditions, we propose \textbf{E}xperience-\textbf{C}alibrated \textbf{C}ontrastive \textbf{D}ecoding (\textbf{ECCD}), a training-free method that strengthens alignment support while preserving useful experience information. ECCD preserves the original expert distribution, applies only positive alignment enhancement, and calibrates its strength using set-level experience compatibility. Across four models, ECCD reduces WER/CER by up to 55.6\% in all SeedTTS-Eval settings and 24 of 25 multilingual CV3-Eval settings. A listening test yields a CMOS gain of $+0.644$ while retaining strong speaker similarity. Further analysis shows that alignment influence and decision-level gain vary within linguistic units and are lower at first-error boundaries than at matched correct boundaries. Overall, these extensive experiments and analyses identify conditional information control as a promising decoding-time direction for mitigating speech hallucination.
\end{abstract}


\section{Introduction}

Neural TTS has evolved from conventional cascaded pipelines \cite{6495700} to natural and expressive end-to-end synthesis \cite{wang2017tacotron,li2019neural}. Combining autoregressive language modeling \cite{yang2025qwen3,grattafiori2024llama} with discrete speech tokenization \cite{zhang2023speechtokenizer} has further yielded LM-based TTS, which predicts acoustic tokens sequentially and supports zero-shot, expressive, and long-form generation \cite{wang2023neural,du2025cosyvoice,hu2026qwen3}. Yet these systems remain vulnerable to speech hallucinations that deviate from the input text, including mispronunciations, substitutions, omissions, repetitions, and unintended continuations \cite{han2024vall}. Such failures are especially pronounced for long, repetitive, or phonetically challenging utterances. 
Existing mitigation mainly focuses on stronger architectures \cite{du2025vallt,song2025ella}, improved training data, or training and post-training objectives \cite{liu2025mitigating,wang2025eliminating}. By contrast, decoding strategies for autoregressive generation remain comparatively underexplored, although they directly control acoustic token selection and thereby affect content faithfulness and speech quality.

Most LM-based TTS systems inherit sampling-based decoding strategies from text generation, such as top-$k$ \cite{fan2018hierarchical} and nucleus sampling \cite{holtzman2019curious}.
Acoustic tokens, however, encode fine-grained, locally correlated speech patterns \cite{tu2024enabling}, leaving several plausible candidates with similar probabilities. These probabilities jointly reflect textual guidance, acoustic history, and learned speech regularities. Conventional sampling does not distinguish these sources and may select a locally plausible but text-inconsistent token. Once added to the autoregressive history, such a token can redirect later predictions and turn a local deviation into a persistent content error. 

To make this ambiguity explicit, we present a conditional information view of speech generation that distinguishes two complementary roles:
\begin{itemize}
    \item \textbf{Alignment information} derives from the text conditions. It
    promotes faithfulness to the target content.
    \item \textbf{Experience information} derives from the acoustic context
    and learned speech regularities. It supports fluent, locally plausible
    speech.
\end{itemize}
This distinction provides a basis for analyzing their effects on token selection and controlling their relative influence during decoding. Both forms are indispensable, and their relative roles vary throughout generation. During the prediction phase, the full-condition distribution reflects both sources, whereas its text-ablated counterpart serves as an operational proxy for experience-oriented prediction.
Notably, hallucinated speech often remains fluent and speech-like after deviating from the target text \cite{liu2025mitigating}. This suggests that useful experience information may remain active while alignment support is not effectively reflected in the decoding decision. We hypothesize that an important class of hallucinations begins with such a shortfall near a vulnerable transition. 
Once an off-target token enters the history, subsequent predictions are conditioned on an already deviated acoustic context. Renewed alignment influence may therefore fail to restore the intended correspondence, allowing the error to propagate.
These dynamics motivate stronger alignment support at vulnerable decisions without sacrificing experience information required for fluency.

Contrastive decoding (CD) \cite{li2023contrastive} provides a natural starting point for this intervention: it selectively amplifies the influence of one, i.e., an expert distribution, by contrasting it against the other, i.e., an amateur reference. 
We instantiate the expert and amateur with the full-condition and text-ablated predictions, respectively. Their token-wise contrast thus provides an operational signal of incremental alignment support.
However, conventional CD treats amateur support solely as negative evidence. This treatment is problematic for speech generation because the proxy retains experience information for pronunciation and temporal continuity. Uniformly suppressing it may damage valid acoustic structure and disrupt the balance required for fluency.

We therefore propose \emph{Experience-Calibrated Contrastive Decoding} (ECCD), a training-free method that selectively strengthens alignment support without suppressing useful experience information. It anchors decoding to the original expert score and applies only positive alignment enhancement within a plausible set defined by the expert. 
We further introduce an \emph{Experience Compatibility Coefficient} (ECC) to scale this enhancement according to the support that the experience proxy assigns to this set.

Across four models, ECCD reduces content errors in all SeedTTS-Eval \cite{anastassiou2024seed} settings and 24 of 25 multilingual CV3-Eval \cite{du2025cosyvoice} settings while preserving perceptual quality and speaker similarity. 
We further validate our conditional information view by measuring how text conditioning reshapes candidate distributions and selected tokens.
These measures show systematic dynamics of both effects within ASR-aligned regions and weakness in the gaps between adjacent regions. Near the first error boundaries, their values fall below those at matched correct boundaries, with a larger shortfall for the selected token. Complete error regions, however, show slightly higher values, suggesting that textual influence may return without restoring the correct mapping. These analysis results provide empirical support for a localized alignment shortfall near onset rather than uniformly weak alignment during propagation.

To the best of our knowledge, this is the first decoding-time conditional-information analysis of LM-based TTS hallucination and the first adaptation of CD to autoregressive acoustic-token generation. Our main contributions are:
\begin{itemize}
    \item We present a conditional information view that distinguishes the
    localized onset pattern of speech hallucination from its autoregressive
    propagation.
    \item We propose ECCD to selectively strengthen alignment support, preserve useful experience information, and dynamically calibrate the enhancement strength.
    \item We provide empirical analyses and evaluations across models,
    datasets, and languages, showing broad decoding-time effectiveness without
    additional training.
\end{itemize}

\section{Related Work}
\subsection{LM-based TTS and Speech Hallucination}
LM-based TTS autoregressively generates discrete codec tokens \cite{wang2023neural}, enabling high-fidelity synthesis, zero-shot speaker transfer, and unified large-scale modeling \cite{kharitonov2023speak,casanova2024xtts,lajszczak2024base}. Recent work improves controllability and structure through unified decoding \cite{song2024touchtts}, decoupled semantic and acoustic modeling \cite{wang2025spark}, and dual-track generation \cite{hu2026qwen3}. Nevertheless, autoregressive systems can produce substitutions, omissions, repetitions, mispronunciations, and unintended continuations.

Previous mitigation primarily strengthens text-to-speech alignment. Guided-attention training combines connectionist temporal classification objectives with attention priors \cite{neekhara2024robustness}; ELLA-V interleaves phoneme and acoustic tokens \cite{song2025ella}; and VALL-T explicitly models monotonic generation with a generative transducer \cite{du2025vallt}. At inference time, attention-constrained inference manipulates alignment-related attention heads \cite{wang2024attention}, while GOAT applies GFlowNet-based post-training to distribution alignment \cite{liu2025mitigating}. These methods largely rely on architectural, training, or model-specific interventions, leaving output-distribution control during decoding underexplored.

\subsection{Decoding-Time Control and Contrastive Decoding}
Practical LM-based TTS systems largely inherit sampling-based decoding strategies, such as top-$k$ \cite{fan2018hierarchical} and nucleus sampling \cite{holtzman2019curious}, which are often combined with temperature scaling and repetition penalties \cite{peng2024voicecraft,chen2024vall,du2024cosyvoice2}. Sampling is efficient but susceptible to stochastic errors that can propagate through long, correlated acoustic token sequences. Greedy and beam search are more deterministic but can exhibit repetitive degeneration and incur greater expansion cost \cite{tu2024enabling,song2025ella}. These trade-offs motivate lightweight decoding-time control.

Contrastive decoding was introduced as a training-free alternative for open-ended text generation \cite{li2023contrastive}. It contrasts a strong expert with a weaker amateur under an expert-defined plausibility constraint. Later work constructs contrasts from predictions with and without external context \cite{shi2024trusting}, different model layers \cite{chuang2024dola}, or different degrees of source grounding \cite{waldendorf2024contrastive}. Related ideas extend to distorted visual inputs \cite{leng2024mitigating} and removed or temporally smoothed audio conditions \cite{hsu2025reducing,li2026temporal}. These studies focus primarily on textual outputs, with limited attention to autoregressive acoustic-token generation.

\section{Method}
\label{sec:method}

\subsection{Problem Formulation}
\label{sec:problem_formulation}

An LM-based text-to-speech (TTS) system autoregressively predicts acoustic tokens before reconstructing the waveform. Given prompt speech $a$, its transcription $T_p$, and target text $T_t$, the LM generates
$\mathbf{x}=(x_1,\ldots,x_N)$ according to
\begin{equation}
    p(\mathbf{x}\mid a,T_p,T_t)
    =
    \prod_{i=1}^{N}p(x_i\mid x_{<i},a,T_p,T_t),
    \label{eq:ar_factorization}
\end{equation}
where $x_{<i}$ denotes the generated history. At each auto-regressive step, the decoder selects $\hat{x}_i$ from the next-token distribution and appends it to the history.



The prediction draws jointly on the text conditions $T_p$ and $T_t$, the acoustic context $(a,x_{<i})$, and the speech prior learned by the LM. These sources support partly distinct aspects of generation, allowing a sequence to remain speech-like even when its content deviates from $T_t$. Such speech-like yet text-inconsistent outputs represent a common form of speech hallucination and motivate our conditional information view.

\subsection{A Conditional Information View of Speech Hallucination}
\label{sec:conditional_information_view}

\paragraph{Dynamic balance between alignment and experience.}
We distinguish two complementary sources of conditional information.
\emph{Alignment information} denotes the text-dependent information supplied by $T_p$ and $T_t$, which establishes the prompt correspondence and guides generation toward the target content. \emph{Experience information} denotes the acoustic continuity, prosody, duration, and locally plausible acoustic and linguistic continuations supported by $a$, $x_{<i}$, and the learned speech prior. This distinction is operational rather than a strict decomposition of the predictive distribution. Both sources jointly shape each next-token prediction.

The relative influence of these sources varies during generation. Near the onset of a linguistic unit, alignment information primarily specifies phonetic content, while experience information maintains acoustic continuity. During sustained or final portions and brief gaps, experience information provides stronger local constraints on pronunciation, duration, and prosody, while alignment information maintains content faithfulness. Their incremental effects also depend on how strongly the generated history already anticipates the target. Experience information may include locally plausible linguistic continuations, not only acoustic regularities. For familiar continuations, the two sources can favor similar candidates; for unusual or difficult content, text conditions may need to reorganize a candidate region favored by the history. Speech generation therefore requires a time-varying balance rather than a fixed relative contribution from either information source.

\paragraph{Hallucination as a dynamic information imbalance.}
Speech hallucination may appear as omissions, repetitions, mispronunciations, substitutions, or unintended continuations. Although the content deviates from the target text, the output often retains fluency, linguistic structure, and prosody \cite{liu2025mitigating}. This suggests that experience information can remain active even when alignment information is not effectively reflected in the decoding decision.

We hypothesize that an important class of hallucinations begins when selected tokens receive insufficient relative alignment support near a vulnerable transition. This shortfall may arise when text induces only a weak distributional change or when a broader change is not sufficiently reflected in the selected token. We call the resulting initial deviation an \emph{onset failure}. Once an erroneous token enters $x_{<i}$, subsequent alignment information interacts with an off-target acoustic history. This produces a text--history conflict in which renewed textual guidance is interpreted through an deviated acoustic context. Its influence may increase without restoring the intended text-to-speech correspondence, allowing the deviation to persist as a \emph{propagation failure}.

\paragraph{Operationalizing alignment and experience information.}
We construct an operational proxy for experience-oriented prediction by retaining the acoustic prompt and generation history while removing both text conditions:
\begin{equation}
    p_A(x_i)\triangleq p(x_i\mid x_{<i},a).
    \label{eq:amateur_distribution}
\end{equation}
The original full-information distribution is
\begin{equation}
    p_E(x_i)\triangleq p(x_i\mid x_{<i},a,T_p,T_t).
    \label{eq:expert_distribution}
\end{equation}
Because $p_E$ and $p_A$ share the same model, acoustic prompt, and history, their discrepancy can be operationally attributed to the incremental influence of text conditions. This comparison is not an exact decomposition of alignment and experience information: $p_A$ is a within-model intervention, and removing text may itself induce a conditioning shift.

We measure distribution-level alignment influence using forward KL divergence:
\begin{equation}
    \mathcal{I}_i
    \triangleq
    D_{\mathrm{KL}}(p_E\parallel p_A)
    =
    \sum_{x_i'\in\mathcal{V}}p_E(x_i')
    \log\frac{p_E(x_i')}{p_A(x_i')},
    \label{eq:alignment_influence}
\end{equation}
where $\mathcal{V}$ is the acoustic-token vocabulary. The direction $D_{\mathrm{KL}}(p_E\parallel p_A)$ takes the expectation under the full-information distribution, and therefore measures how strongly its candidate distribution differs from the text-ablated reference. To determine whether this distributional influence is reflected in the realized decision, we define the decision-level alignment gain for the selected token $\hat{x}_i$ as
\begin{equation}
    \mathcal{G}_i
    \triangleq
    \log p_E(\hat{x}_i)-\log p_A(\hat{x}_i).
    \label{eq:selected_token_gain}
\end{equation}
Under our operationalization, a positive $\mathcal{G}_i$ means that full conditioning increases the selected token's relative likelihood over text ablation, whereas a non-positive value indicates no positive relative alignment support.
To make the relationship between the two measures explicit, define $g_i(x_i')=\log p_E(x_i')-\log p_A(x_i')$. 
We then have $\mathcal{I}_i=\mathbb{E}_{x_i'\sim p_E}[g_i(x_i')]$, whereas $\mathcal{G}_i=g_i(\hat{x}_i)$. Thus, the two measures evaluate the same alignment-induced contrast at the distribution and decision levels, respectively. A large distribution-level influence does not necessarily imply strong alignment support for the token ultimately selected.

Our hypothesis concerns a local shortfall in $\mathcal{G}_i$ near first-error onset, not uniformly weak alignment throughout an error span. Neither measure is a correctness score, and both may increase after the history deviates without indicating recovery. We examine onset windows, complete regions, and relative-position profiles separately in the analysis below.

\subsection{Experience-Calibrated Contrastive Decoding}
\label{sec:eccd}

\paragraph{Conventional contrastive decoding.}



The preceding information view suggests that an important class of hallucination onset may be mitigated by strengthening decision-level alignment support while preserving the experience information required for natural speech. Contrastive decoding (CD) \cite{li2023contrastive} provides a natural starting point. With $p_E$ as the expert and $p_A$ as the amateur, their token-wise log-likelihood ratio $\log p_E(x_i)-\log p_A(x_i)$ serves as an operational contrast for alignment support. Because this ratio can become large even when $p_E(x_i)$ is negligible, we restrict CD to the top-$k$ candidates under $p_E$:
\begin{equation}
    \mathcal{V}_{\mathrm{head}}(x_{<i})
    \triangleq
    \operatorname{TopK}_{k}
    \left(\{p_E(x_i)\}_{x_i\in\mathcal{V}}\right).
    \label{eq:plausible_set}
\end{equation}
The set is recomputed at every decoding step. It makes the admissible range explicit and prevents a large relative contrast from promoting candidates that lack sufficient support under full conditioning. The resulting conventional CD score is
\begin{equation}
    \mathcal{S}_{\mathrm{CD}}(x_i)
    =
    \begin{cases}
        \displaystyle \log\dfrac{p_E(x_i)}{p_A(x_i)},
        & x_i\in\mathcal{V}_{\mathrm{head}},\\
        -\infty,
        & x_i\notin\mathcal{V}_{\mathrm{head}},
    \end{cases}
    \label{eq:vanilla_cd_score}
\end{equation}
and $p_{\mathrm{CD}}=\operatorname{Softmax}(\mathcal{S}_{\mathrm{CD}})$.

Conventional CD treats $p_A$ solely as negative evidence. Within the plausible set, the likelihood ratio becomes the complete decoding score rather than a correction to $p_E$, thus amateur support contributes only subtractively. Outside the set, the original expert scores are discarded. This treatment is problematic for speech because $p_A$ retains useful pronunciation, prosody, duration, and continuity information. Moreover, a fixed contrastive transformation cannot reflect the time-varying relation between alignment and experience information. 
These limitations motivate our Experience-Calibrated Contrastive Decoding (ECCD).

\paragraph{ECCD formulation.}
ECCD addresses these limitations through \emph{experience preservation} and
\emph{experience calibration}. Its next-token score is
\begin{equation}
    \mathcal{S}_{\mathrm{ECCD}}(x_i)
    =
    \begin{cases}
        \begin{aligned}
            &\log p_E(x_i)+\alpha(1-\mathcal{C}_i)\\
            &\quad\cdot
            \left[\log\dfrac{p_E(x_i)}{p_A(x_i)}\right]_+
        \end{aligned}
        & x_i\in\mathcal{V}_{\mathrm{head}},\\
        \log p_E(x_i)
        & x_i\notin\mathcal{V}_{\mathrm{head}}.
    \end{cases},
    \label{eq:eccd_score}
\end{equation}
where $\alpha>0$ controls enhancement strength and $[\,\cdot\,]_+\triangleq\max(0,\,\cdot\,)$. The Experience Compatibility Coefficient (ECC) is
\begin{equation}
    \mathcal{C}_i
    \triangleq
    \sum_{x_i'\in\mathcal{V}_{\mathrm{head}}(x_{<i})}p_A(x_i'),
    \qquad 0\leq\mathcal{C}_i\leq1.
    \label{eq:experience_compatibility}
\end{equation}
Here, $k$ controls the plausible-set size. The decoding distribution is thus $p_{\mathrm{ECCD}}=\operatorname{Softmax}(\mathcal{S}_{\mathrm{ECCD}})$. The score in Eq.~(\ref{eq:eccd_score}) combines three components: the expert anchor $\log p_E(x_i)$, the positive-only enhancement $[\log p_E(x_i)/p_A(x_i)]_+$, and the calibration factor $(1-\mathcal{C}_i)$. We now explain how the first two preserve experience information and how the third adapts enhancement strength.

\paragraph{Experience preservation.}
The full-information distribution already combines alignment and experience information. ECCD therefore retains $\log p_E(x_i)$ as an expert anchor, thus the alignment contrast modifies rather than replaces this integrated score. The plausible set controls where the correction is applied rather than which tokens remain available; candidates outside $\mathcal{V}_{\mathrm{head}}$ keep their expert scores.

Within the plausible set, ECCD promotes only candidates with $p_E(x_i)>p_A(x_i)$. Candidates preferred by $p_A$ are not explicitly penalized because their support may reflect useful pronunciation, duration, or continuity patterns. Expert anchoring and positive-only enhancement thus increase operational alignment support without directly treating experience information as an error signal.

\paragraph{Experience calibration.}
Experience preservation specifies how the alignment contrast modifies the full-information score, but not how strongly it should be applied. A fixed strength is unsuitable because the relative roles of alignment and experience information vary during generation. We therefore use ECC $\mathcal{C}_i$ to measure how much probability mass the experience proxy assigns to the plausible set.

When generated history and learned speech regularities support candidates similar to those favored under full conditioning, $\mathcal{C}_i$ is large. The original distribution then reflects a compatible integration of both sources. Further enhancement may unnecessarily shift probability mass toward alignment-supported candidates and away from experience-supported duration and continuity patterns. ECCD therefore attenuates the intervention to preserve the existing information balance and reduce possible temporal compression.



Conversely, a small $\mathcal{C}_i$ indicates limited experience support for the expert-defined plausible set and hence a greater discrepancy between the two predictions. At such a step, failure to translate the alignment-induced distributional change into support for the selected token may increase vulnerability to hallucination onset. ECCD therefore strengthens the positive correction for expert-plausible candidates whose likelihood increases under full conditioning, encouraging distribution-level alignment influence to be more effectively reflected in token selection. Here, $\mathcal{C}_i$ is a bounded set-level compatibility proxy for calibration. It is neither a detector of hallucination onset nor a replacement for the distribution-level measure $\mathcal{I}_i$.

\begin{table*}[t]
    \centering
    \small
    \setlength{\tabcolsep}{4pt}
    \begin{tabular}{@{}lccccccccc@{}}
        \toprule
        \multirow{2}{*}{Model}
        & \multicolumn{3}{c}{\textit{test-zh}}
        & \multicolumn{3}{c}{\textit{test-en}}
        & \multicolumn{3}{c}{\textit{test-hard}} \\
        \cmidrule(lr){2-4}
        \cmidrule(lr){5-7}
        \cmidrule(lr){8-10}
        & CER (\%) $\downarrow$ & SS $\uparrow$ & UTMOS $\uparrow$
        & WER (\%) $\downarrow$ & SS $\uparrow$ & UTMOS $\uparrow$
        & CER (\%) $\downarrow$ & SS $\uparrow$ & UTMOS $\uparrow$ \\
        \midrule

        Human
        & 1.31 & 0.776 & 2.783 $\pm$ 0.021
        & 2.74 & 0.820 & 3.535 $\pm$ 0.031
        & -- & -- & -- \\

        \midrule
        CosyVoice2
        & 1.34 & 0.849 & 3.474 $\pm$ 0.018
        & 2.98 & \textbf{0.810} & 4.153 $\pm$ 0.018
        & 8.10 & \textbf{0.822} & 3.313 $\pm$ 0.045 \\

        \quad \quad +LT ($\tau=0.75$)
        & 1.11 & \textbf{0.850} & \textbf{3.517 $\pm$ 0.018}
        & 4.35 & 0.799 & \textbf{4.190 $\pm$ 0.018}
        & 7.28 & 0.819 & 3.376 $\pm$ 0.044 \\

        \quad \quad +LT ($\tau=0.875$)
        & 1.18 & 0.848 & 3.500 $\pm$ 0.018
        & 2.76 & 0.807 & 4.173 $\pm$ 0.018
        & 7.60 & 0.819 & \textbf{3.388 $\pm$ 0.045} \\

        \quad \quad +ECCD
        & \textbf{0.95} & 0.841 & 3.453 $\pm$ 0.018
        & \textbf{2.08} & \textbf{0.810} & 4.176 $\pm$ 0.017
        & \textbf{6.91} & 0.811 & 3.265 $\pm$ 0.044 \\

        \midrule
        CosyVoice3
        & 1.19 & \textbf{0.876} & 3.317 $\pm$ 0.018
        & 3.01 & \textbf{0.819} & 3.941 $\pm$ 0.023
        & 6.92 & \textbf{0.859} & \textbf{3.172 $\pm$ 0.043} \\

        \quad \quad +ECCD
        & \textbf{0.92} & 0.867 & \textbf{3.318 $\pm$ 0.018}
        & \textbf{1.82} & 0.818 & \textbf{3.977 $\pm$ 0.021}
        & \textbf{6.19} & 0.846 & 3.117 $\pm$ 0.043 \\

        \midrule
        Llasa
        & 8.66 & 0.651 & 3.252 $\pm$ 0.029
        & 6.83 & 0.744 & 4.018 $\pm$ 0.031
        & 26.26 & 0.579 & 2.934 $\pm$ 0.074 \\

        \quad \quad +ECCD
        & \textbf{3.95} & \textbf{0.675} & \textbf{3.352 $\pm$ 0.026}
        & \textbf{3.03} & \textbf{0.750} & \textbf{4.106 $\pm$ 0.025}
        & \textbf{14.27} & \textbf{0.609} & \textbf{3.060 $\pm$ 0.067} \\

        \midrule
        GLM-TTS
        & 1.01 & \textbf{0.867} & \textbf{2.935 $\pm$ 0.021}
        & 2.12 & \textbf{0.826} & 3.704 $\pm$ 0.030
        & 11.11 & \textbf{0.858} & \textbf{2.687 $\pm$ 0.048} \\

        \quad \quad +ECCD
        & \textbf{0.97} & 0.854 & 2.923 $\pm$ 0.020
        & \textbf{1.94} & 0.825 & \textbf{3.709 $\pm$ 0.029}
        & \textbf{9.70} & 0.839 & 2.604 $\pm$ 0.046 \\

        \bottomrule
    \end{tabular}
    \caption{SeedTTS-Eval objective results. WER and CER are percentages; UTMOS is
    reported as mean $\pm$ 95\% confidence-interval half-width; LT denotes
    lower-temperature sampling.}
    \label{tab:seedtts_objective}
\end{table*}

\section{Experiments}

\subsection{Experimental Setup}
\label{sec:experimental_setup}

\paragraph{Models, datasets, and baselines.}
We evaluate ECCD with CosyVoice2/3 \cite{du2024cosyvoice2, du2025cosyvoice}, Llasa \cite{ye2025llasa} and GLM-TTS \cite{cui2025glm}. All models are tested on the \textit{test-en}, \textit{test-zh}, and \textit{test-hard} subsets of SeedTTS-Eval \cite{anastassiou2024seed}. We further evaluate each model on its supported languages in the zero-shot track of CV3-Eval \cite{du2025cosyvoice}, covering nine languages. In addition to native decoding, lower-temperature sampling with $\tau\in\{0.75,0.875\}$ tests whether ECCD merely benefits from distribution sharpening.

\paragraph{Implementation details.}
The expert and amateur share model parameters and the same realized history; the expert receives all model-specific conditions, while the amateur has its text conditions ablated. At every step, the realized history is teacher-forced into both branches, hence their predictions differ operationally in the availability of text conditions. ECCD is training-free and modifies next-token scores before the model's native filtering and penalty operations. Unless stated otherwise, $\alpha=1$ and $k=25$. All inference and objective evaluations run on an NVIDIA RTX 3090 GPU. 

\paragraph{Evaluation metrics.}
We report Word/Character Error Rate (W/CER) as content-fidelity metrics and operational proxies for hallucination severity. Paraformer-zh \cite{gao2022paraformer} transcribes Chinese speech, and Whisper-large-v3 \cite{radford2023robust} handles the remaining languages. Speaker similarity (SS) uses CAM++ \cite{wang2023cam++}, and UTMOS \cite{saeki2022utmos} estimates speech quality. We also conduct a listening test on CosyVoice2 using \textit{test-hard}. Twenty five native listeners evaluate 30 utterances per method: CMOS compares content-aware preference against native decoding on a $-3$ to $3$ scale, with target text shown, while SMOS measures prompt-based speaker similarity on a $1$ to $5$ scale. The mechanism analysis uses $\mathcal{I}_i$ and $\mathcal{G}_i$ as defined above.

\subsection{Main Results}
\label{sec:main_results}

\paragraph{Objective evaluation.}
Table~\ref{tab:seedtts_objective} shows that ECCD consistently reduces WER/CER across all four base models and every SeedTTS-Eval subset. The improvement is most pronounced for Llasa: CER decreases from 8.66\% to 3.95\% on \textit{test-zh} and from 26.26\% to 14.27\% on \textit{test-hard}, while WER on \textit{test-en} falls from 6.83\% to 3.03\%.


We further compare ECCD with lower-temperature sampling on CosyVoice2. Although this baseline improves several settings, its performance is inconsistent. At $\tau=0.75$, \textit{test-en} WER increases from 2.98\% to 4.35\%. By contrast, ECCD attains the lowest WER/CER across all three subsets, suggesting that its gains are not reproduced by distribution sharpening alone. For CosyVoice and GLM-TTS, the maximum decreases in SS and UTMOS are 0.019 and 0.083, whereas both metrics improve for Llasa. The CosyVoice shifts may partly reflect altered temporal progression, which the ablation below quantifies. 
We also report listener ratings because automatic metrics may not reflect human judgments.

\paragraph{Multilingual generalization.}
\begin{table}[t]
    \centering
    \small
    \setlength{\tabcolsep}{7pt}
    \begin{tabular}{@{}lcccc@{}}
        \toprule
        Model
        & \textit{zh}
        & \textit{hard-zh}
        & \textit{en}
        & \textit{hard-en} \\
        \midrule

        CosyVoice2
        & 4.08 & 13.21 & 6.22 & 10.18 \\
        \quad \quad +ECCD
        & \textbf{3.58} & \textbf{12.68}
        & \textbf{4.90} & \textbf{8.14} \\

        \midrule
        CosyVoice3
        & 3.90 & 8.89 & 5.66 & 8.84 \\
        \quad \quad +ECCD
        & \textbf{3.34} & \textbf{8.38}
        & \textbf{4.16} & \textbf{7.79} \\

        \midrule
        Llasa
        & 18.24 & 34.38 & 26.31 & 39.20 \\
        \quad \quad +ECCD
        & \textbf{13.40} & \textbf{22.17}
        & \textbf{20.35} & \textbf{24.16} \\

        \midrule
        GLM-TTS
        & 3.68 & 9.31 & 6.54 & 7.88 \\
        \quad \quad +ECCD
        & \textbf{3.42} & \textbf{8.49}
        & \textbf{5.66} & \textbf{5.63} \\

        \bottomrule
    \end{tabular}
    \caption{Zero-shot CV3-Eval WER/CER (\%) for Chinese and
    English subsets.}
    \label{tab:cv3_zh_en}
\end{table}

\begin{table}[t]
    \centering
    \small
    \setlength{\tabcolsep}{4pt}
    \begin{tabular}{@{}lccccccc@{}}
        \toprule
        Model
        & \textit{ja}
        & \textit{ko}
        & \textit{de}
        & \textit{es}
        & \textit{fr}
        & \textit{it}
        & \textit{ru} \\
        \midrule

        CosyVoice2
        & \textbf{9.46} & 6.50
        & -- & -- & -- & -- & -- \\
        \quad \quad +ECCD
        & 9.71 & \textbf{5.39}
        & -- & -- & -- & -- & -- \\

        \midrule
        CosyVoice3
        & 12.13 & 6.56 & 6.87 & 4.35
        & 10.99 & 6.77 & 7.40 \\
        \quad \quad +ECCD
        & \textbf{10.29} & \textbf{4.96}
        & \textbf{5.65} & \textbf{3.76}
        & \textbf{10.07} & \textbf{4.95}
        & \textbf{5.45} \\

        \bottomrule
    \end{tabular}
    \caption{Zero-shot CV3-Eval WER/CER (\%) for additional
    languages. Dashes denote unsupported languages.}
    \label{tab:cv3_additional_languages}
\end{table}

Table~\ref{tab:cv3_zh_en} and Table~\ref{tab:cv3_additional_languages} show that ECCD reduces WER/CER in 24 of 25 supported model and subset combinations, including every \textit{hard-zh} and \textit{hard-en} setting. Gains extend to Korean, German, Spanish, French, Italian, and Russian without language-specific training, supporting broad cross-lingual generalization over the evaluated models. This training-free transfer is particularly useful when language-specific data or post-training resources are limited.

The only exception is Japanese with CosyVoice2, where WER/CER rises slightly from 9.46\% to 9.71\%, while falling from 12.13\% to 10.29\% with CosyVoice3. The CosyVoice3 study relates Japanese difficulties in CosyVoice2 to overlapping Chinese and Japanese characters, and reports converting Japanese text to kana \cite{du2025cosyvoice}. We therefore hypothesize that this isolated reversal reflects a base-model text-representation ambiguity that decoding-time alignment enhancement cannot remove, although dedicated analysis would be required to establish the cause.

\paragraph{Subjective evaluation.}
\label{sec:subjective_evaluation}
\begin{table}[t]
    \centering
    \small
    \setlength{\tabcolsep}{7pt}
    \begin{tabular}{@{}lcc@{}}
        \toprule
        Method & CMOS $\uparrow$ & SMOS $\uparrow$ \\
        \midrule
        CosyVoice2
        & 0.000
        & 3.678 $\pm$ 0.103 \\

        \quad \, +LT ($\tau=0.75$)
        & +0.021 $\pm$ 0.086
        & 3.831 $\pm$ 0.129 \\

        \quad \, +LT ($\tau=0.875$)
        & +0.037 $\pm$ 0.082
        & \textbf{3.962 $\pm$ 0.116} \\

        \quad \, +ECCD
        & \textbf{+0.644 $\pm$ 0.091}
        & 3.811 $\pm$ 0.113 \\
        \bottomrule
    \end{tabular}
    \caption{CosyVoice2 subjective results on SeedTTS-Eval
    \textit{test-hard} (mean $\pm$ 95\% confidence-interval half-width; CMOS
    relative to native decoding).}
    \label{tab:subjective_evaluation}
\end{table}

Since \textit{test-hard} lacks reference recordings, CMOS compares each method with native CosyVoice2, whereas SMOS uses prompt speech as the speaker reference. Listeners receive the target text, allowing CMOS to reflect perceptual quality and content faithfulness.

Table~\ref{tab:subjective_evaluation} shows a CMOS gain of $+0.644$ for ECCD, versus $+0.021$ and $+0.037$ for lower-temperature sampling, consistent with its larger reduction in hallucination-related errors. ECCD's SMOS of $3.811$ is below the lower-temperature variants but above the native score of $3.678$. Overall, the objective and listening results show content-fidelity gains with modest quality and speaker-similarity trade-offs.

\subsection{Information-Theoretic Analysis of Hallucination}
\label{sec:hallucination_analysis}
We next examine the temporal conditional-information patterns defined above in relation to hallucination onset.

\paragraph{Analysis setup.}
We analyze native CosyVoice2 outputs on SeedTTS-Eval \textit{test-hard}. Character-level correct/error labels are obtained by aligning ASR transcriptions with target text, and acoustic-token spans are approximated using ASR timestamps and the model token rate; gaps provide a temporal reference. Within each character region, offset $0$ is its first mapped token, offset $-1$ is the preceding token, and positive offsets denote later tokens.

For onset analysis, we use the first time-localizable error in each erroneous utterance and the immediately preceding correct region as its matched reference. All statistics are computed from native-decoder outputs independently of ECCD, and token-level measurements are pooled across matched windows as arithmetic means. The pre-boundary window $\{-1\}$ probes the state immediately before the observed boundary, whereas $\{-1,0\}$ straddles it and is less sensitive to timestamp quantization. Since ASR labels, timestamps, and token-rate mapping are approximate, these windows localize the observed transition rather than its exact causal decoding step. 

\begin{table}[t]
    \centering
    \small
    \setlength{\tabcolsep}{10pt}
    \begin{tabular}{@{}lccc@{}}
        \toprule
        Metric & Correct & Error & Gap \\
        \midrule
        $\mathcal{I}_i$ & 0.885 & 0.934 & 0.432 \\
        $\mathcal{G}_i$ & 1.095 & 1.130 & 0.480 \\
        \bottomrule
    \end{tabular}
    \caption{Complete-region means of distribution-level alignment influence
    $\mathcal{I}_i$ and decision-level alignment gain $\mathcal{G}_i$.
    Gaps provide a temporal reference.}
    \label{tab:complete_region_averages}
\end{table}

\begin{table}[t]
    \centering
    \small
    \setlength{\tabcolsep}{4pt}
    \begin{tabular}{@{}lcccc@{}}
        \toprule
        Metric & Window &
        Matched correct &
        First error &
        $\Delta_{\mathrm{C-E}}$ \\
        \midrule
        $\mathcal{I}_i$ & $\{-1\}$   & 0.949 & 0.876 & 0.073 \\
         & $\{-1,0\}$ & 1.029 & 0.970 & 0.059 \\
        \midrule
        $\mathcal{G}_i$ & $\{-1\}$   & 1.244 & 1.007 & 0.237 \\
         & $\{-1,0\}$ & 1.351 & 1.198 & 0.153 \\
        \bottomrule
    \end{tabular}
    \caption{Matched statistics around the first ASR-detected error boundary.
    Correct denotes the preceding correct region, and
    $\Delta_{\mathrm{C-E}}$ is the matched correct-minus-error difference.}
    \label{tab:first_error_onset}
\end{table}




\begin{figure}[t]
\centering
\includegraphics[width=0.95\columnwidth]{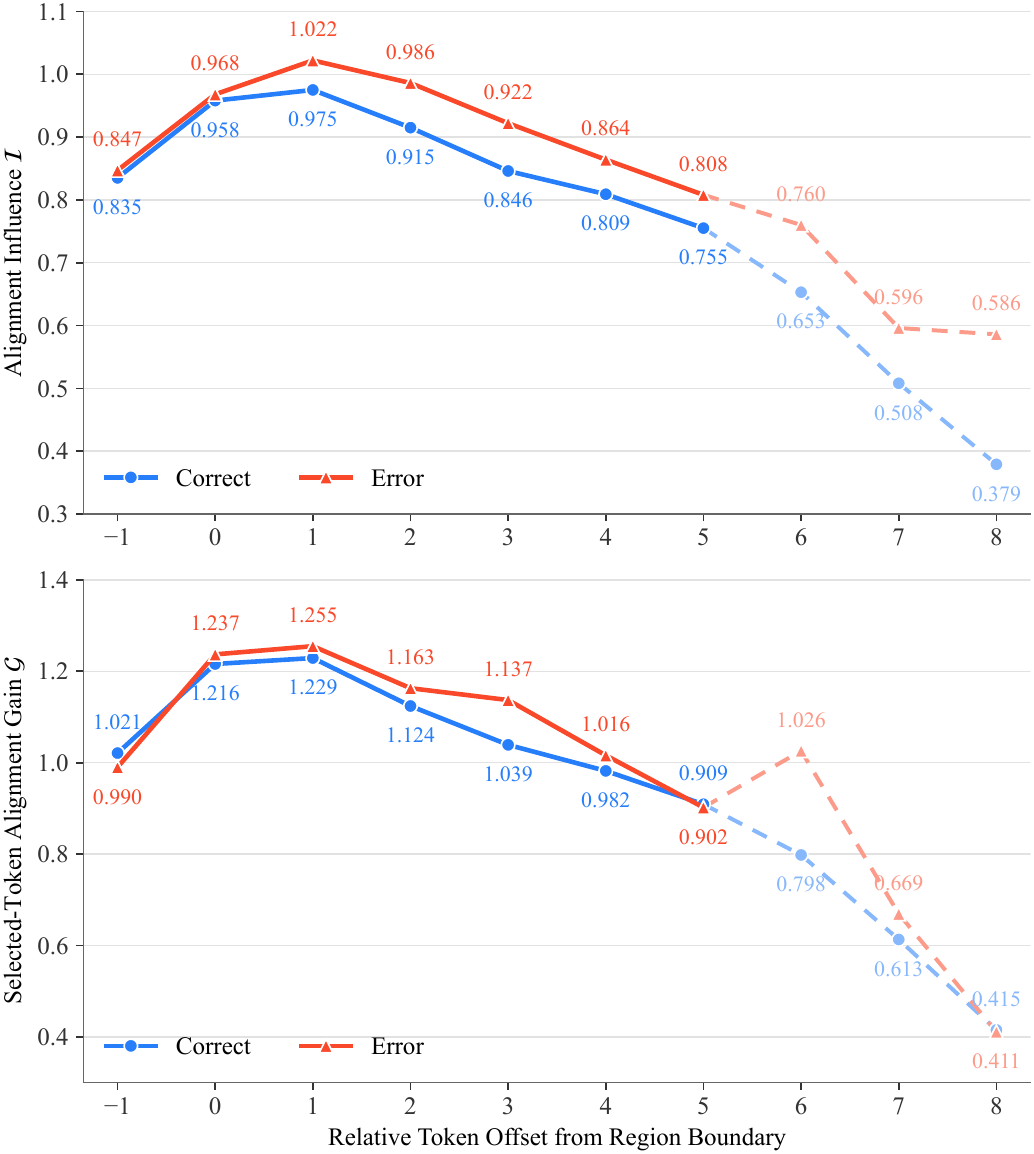} 
\caption{Mean relative-position profiles of $\mathcal{I}_i$ (top) and
$\mathcal{G}_i$ (bottom) over ASR-aligned correct and erroneous character
regions. Offset $0$ is the first token. Segments after offset $5$ are
dashed because their sample counts decrease sharply.}
\label{fig:alignment_relative_profiles}
\end{figure}

\paragraph{Dynamic balance between alignment and experience.}
Figure~\ref{fig:alignment_relative_profiles} shows a common temporal pattern for correct and erroneous regions: $\mathcal{I}_i$ and $\mathcal{G}_i$ rise near character boundaries, peak around offset $1$, and decline during within-region continuation. Table~\ref{tab:complete_region_averages} provides a complementary reference. Gap means, $0.432$ for $\mathcal{I}_i$ and $0.480$ for $\mathcal{G}_i$, are less than half of the corresponding complete-character averages. These patterns are consistent with alignment playing a stronger role near linguistic transitions and experience providing stronger local constraints during sustained pronunciation and gaps. Dashed late-offset segments are descriptive only because their sample counts decrease sharply.

\paragraph{Alignment support at hallucination onset.}
Complete error regions may include propagated errors conditioned on an off-target history. We therefore compare each first-error boundary with its matched correct boundary in Table~\ref{tab:first_error_onset}.
The first-error boundary has lower $\mathcal{I}_i$ and $\mathcal{G}_i$ in both windows. Correct-minus-error differences are $0.073$ and $0.237$ over $\{-1\}$, and $0.059$ and $0.153$ over $\{-1,0\}$, respectively. The larger correct-minus-error gap for $\mathcal{G}_i$ than for $\mathcal{I}_i$ shows that the shortfall is more pronounced at the realized decision than across the complete candidate distribution. The consistently lower values across both matched windows are consistent with, but do not causally establish, an alignment-support shortfall near an important class of hallucination onsets.



After onset, alignment information interacts with an off-target acoustic history. The slightly higher complete-error averages and post-boundary profiles reflect this propagation regime, although intrinsic content difficulty may also contribute. Since neither measure is a correctness score, their increase does not necessarily indicate recovery of the intended correspondence. Future work should investigate how content difficulty and off-target acoustic history jointly shape longer-range information dynamics.

\subsection{Ablation Studies}
\label{sec:ablation}

\begin{table}[t]
    \centering
    \small
    \setlength{\tabcolsep}{4pt}
    \begin{tabular}{@{}lcccc@{}}
        \toprule
        Variant
        & CER$\downarrow$
        & SS$\uparrow$
        & UAC (ratio)
        & TAD (ratio) \\
        \midrule
        Native decoder
        & 8.10
        & \textbf{0.822}
        & 390.24 (1.00)
        & 0.218 (1.00) \\

        Conventional CD
        & 8.72
        & 0.738
        & 256.78 (0.66)
        & 0.166 (0.76) \\

        \quad +Expert anchor
        & 6.86
        & 0.793
        & 302.02 (0.77)
        & 0.189 (0.87) \\

        \quad +Positive-only
        & \textbf{6.43}
        & 0.795
        & 303.89 (0.78)
        & 0.190 (0.87) \\
        \midrule
        \quad +$\mathcal{C}_i$ (ECCD)
        & 6.91
        & 0.811
        & 339.17 (0.87)
        & 0.202 (0.93) \\
        
        \quad +$\mathcal{I}_i$ ($\mathrm{ECCD}_{\mathcal I}$)
        & 6.90
        & 0.799
        & 314.39 (0.81)
        & 0.191 (0.88) \\
        \bottomrule
    \end{tabular}
    \caption{CosyVoice2 component ablation on SeedTTS-Eval
    \textit{test-hard}. Final rows compare calibration signals; parentheses give
    ratios to native decoding, with values closer to $1$ indicating better
    temporal preservation.}
    \label{tab:eccd_component_ablation}
\end{table}

\paragraph{Experience preservation and calibration.}
Table~\ref{tab:eccd_component_ablation} progressively adds expert anchoring, positive-only enhancement, and $\mathcal{C}_i$ to conventional CD.
For $\mathrm{ECCD}_{\mathcal I}$, we replace $1-\mathcal{C}_i$ in Eq.~\ref{eq:eccd_score} with $\mathcal{I}_i$ to compare $\mathcal{I}_i$ as an alternative calibration signal. To quantify temporal compression, we report two descriptive statistics: utterance-level average acoustic-token count (UAC) and timestamp-derived average character duration (TAD).

Conventional CD increases CER from 8.10\% to 8.72\%, reduces SS to 0.738, and compresses UAC and TAD to 66\% and 76\% of native decoding. This behavior is consistent with treating experience-supported preferences solely as negative evidence. Expert anchoring reverses the CER degradation and partly restores temporal structure. Positive-only enhancement further lowers CER to 6.43\% without additional compression, supporting the importance of avoiding explicit penalties on experience-supported candidates.

Adding $\mathcal{C}_i$ trades part of the maximum CER reduction for stronger preservation. Relative to the positive-only variant, it raises SS from 0.795 to 0.811 and restores the UAC/TAD ratios from 0.78/0.87 to 0.87/0.93. The larger UAC recovery is consistent with restoring tokens outside ASR-aligned character spans, including gaps, although these aggregate measures cannot localize them exactly. Using $\mathcal{I}_i$ as the calibration signal gives nearly identical CER but lower SS and UAC/TAD ratios, suggesting that set-level experience compatibility better matches the calibration objective in this setting.

\paragraph{Hyperparameter sensitivity.}



\begin{figure}[t]
\centering
\includegraphics[width=0.98\columnwidth]{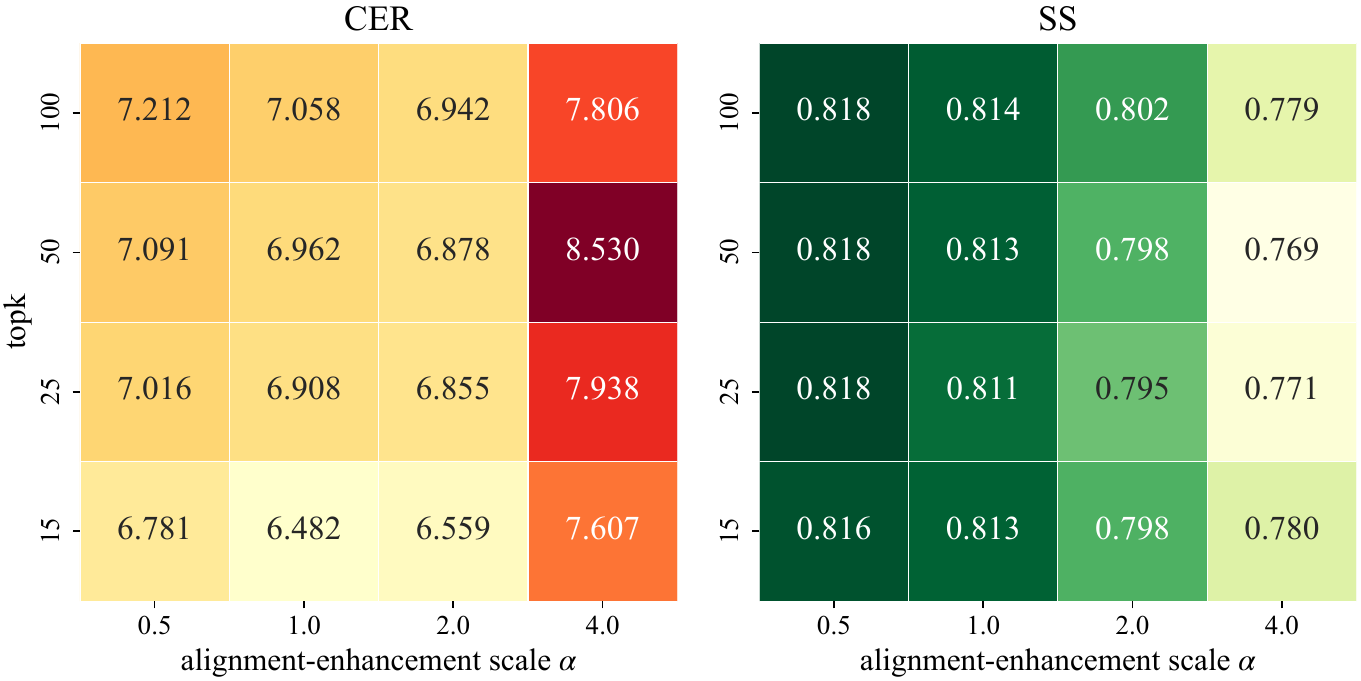} 
\caption{Hyperparameter sensitivity of ECCD on CosyVoice2 using
    SeedTTS-Eval \textit{test-hard}. We vary the alignment-enhancement
    scale $\alpha$ and plausible-set size $k$.}
\label{fig:hyper_sensitivity}
\end{figure}


Figure~\ref{fig:hyper_sensitivity} shows that moderate increases in $\alpha$ generally reduce CER while gradually lowering SS, whereas $k$ has a smaller effect on SS and smaller plausible sets tend to yield lower CER. Setting $\alpha=4$ degrades both metrics across the tested values of $k$, indicating that excessive alignment enhancement disrupts the information balance. Although we use $\alpha=1,k=25$ as a fixed default across experiments, $\alpha=1,k=15$ improves both CER and SS in this sweep, suggesting a favorable deployment setting when validation-based tuning is available.

\section{Conclusion}

We presented ECCD, a training-free decoding method that mitigates hallucination-related content errors by contrasting full-condition and text-ablated predictions. ECCD strengthens alignment support while preserving experience information and calibrates the enhancement strength through set-level compatibility. Objective evaluations across four models show that ECCD reduces WER/CER in all SeedTTS-Eval settings and 24 of 25 multilingual CV3-Eval settings. A listening test complements these results by revealing a positive content-aware preference for ECCD. At the component level, ablations indicate that experience preservation and calibration alleviate the temporal compression caused by conventional CD.
Our empirical analysis further shows that alignment influence and decision-level gain rise near ASR-aligned character boundaries and decline during continuation. Both measures are lower at first-error boundaries than at matched correct boundaries, consistent with an alignment-support shortfall near hallucination onset. 
Overall, these results position conditional-information control as a promising decoding-time complement to training-based methods for mitigating speech hallucinations. 

\bibliography{aaai2027}


\end{document}